\documentclass{PoS}

\PoS{PoS(HEP2005)334}

\title{Searching for new physics in bottomonium decays}

\ShortTitle{Searching for new physics in bottomonium decays}

\author{\speaker{Miguel Sanchis-Lozano}\thanks{Research under grant 
FPA2002-00612, GV-GRUPOS03/094 and GV05/276.}\\
        IFIC-Departamento de Fisica Teorica, Valencia University, Spain\\
        E-mail: \email{Miguel.Angel.Sanchis@uv.es}}

\abstract{Heavy quarkonium decays can be used to search for New Physics
beyond the Standard Model. In particular, a light Higgs boson could induce
a slight (but observable) lepton universality breaking in Upsilon leptonic
decays. In fact, current experimental data from CLEO presented in
this Conference seem to point out to this
direction within experimental accuracy. Moreover, LEP constraints 
on a light Higgs mass can be
evaded by different models (like MSSM with a CPV Higgs sector)
as shown in this Conference. We also
consider spectroscopic consequences stemming from a possible mixing
between Higgs and bottomonium states leading to discrepancies
with the SM expectations (e.g. hyperfine splittings).}

\FullConference{International Europhysics Conference on High Energy Physics\\
		 July 21st - 27th 2005\\
		 Lisboa, Portugal}

\begin{document}

\section{Introduction}
The CLEO collaboration has recently released experimental results
on muonic branching fractions (BF's) of Upsilon resonances below open
bottom threshold \cite{Adams:2004xa}. Moreover, CLEO
has presented in this Conference new and more precise results on the 
tauonic BF's of
all three $\Upsilon(1S)$, $\Upsilon(2S)$ and $\Upsilon(3S)$ states,
the latter representing the first ever observed tauonic
decay rate of this resonance. By combining these 
new data with the already available PDG values
\cite{pdg}, we show in Table 1 the current (though preliminary) 
status of all six BF's, thereby allowing a
test of lepton universality in Upsilon decays. Deviation from this
hypothesis can be assessed using the ratio 
\[
R_{\tau/\ell}(nS)=\frac{\mathcal{B}[\Upsilon(nS) \to \tau
\tau]-\mathcal{B}[\Upsilon(nS) \to \ell
    \ell]}{\mathcal{B}[\Upsilon(nS) \to \ell \ell]}\ \ \ \ ,\ \ \ \ 
\ell=e,\mu
\]
As pointed out in the literature \cite{Sanchis-Lozano:2005fj,
Brambilla:2004,Sanchis-Lozano:2004gh,Sanchis-Lozano:2003ha,
Sanchis-Lozano:2002pm}, 
lepton universality breaking in Upsilon decays would open
up the possibility of New Physics (NP) beyond the Standard Model
(SM), indicating the existence of a light 
non-standard Higgs boson. Such a particle would mediate
the $b\bar{b}$ annihilation into a tauonic pair subsequent to a
dipole magnetic (either allowed or hindered) 
transition of the Upsilon, yielding a soft (unobserved)
photon according to the cascade process
\begin{equation}
\Upsilon(nS) \to \gamma_s\eta_b(n'S)(\to \tau^+\tau^-)\ \ \ \ ;\ \ \ \
n \geq n'=1,2,3
\end{equation}
In our later development,
we shall factorize the above
cascade decay as
\begin{equation}
\mathcal{B}[\Upsilon(nS) \to \gamma_s\tau^+\tau^-] =
\mathcal{B}[\Upsilon(nS) \to \gamma_s\eta_b(n'S)]
\times  \mathcal{B}[\eta_b(n'S) \to \tau^+\tau^-]
\end{equation}
Because the photon would escape unnoticed \footnote{As far as it is
not searched for; it should be possible, however, to look for such soft
photons in the tauonic sub-sample of CLEO on-tape recorded
events}, such a NP contribution would be unwittingly ascribed to
the tauonic decay mode, while the electronic or muonic modes would
result unaltered, ultimately implying lepton
universality breaking.

From a theoretical viewpoint, the existence of a light
pseudoscalar in the Higgs sector is compatible with
certain extensions of the SM involving two Higgs doublets \cite{gunion},
e.g. a Peccei-Quinn symmetry can yield
a pseudo-Nambu-Goldstone boson, which for a range of model
parameters (for example, in the NMSSM) is significantly lighter than
the other scalars.

\begin{table*}[hbt]
\setlength{\tabcolsep}{0.4pc}
\caption{Measured leptonic branching fractions 
${\mathcal{B}[\Upsilon(nS) \to \ell \ell]}$ (in $\%$) and error bars 
(summed in quadrature) of
$\Upsilon(1S)$, $\Upsilon(2S)$ and $\Upsilon(3S)$ resonances 
(obtained from CLEO data given at this Conference and 
\cite{pdg}).}
\label{FACTORES}

\begin{center}
\begin{tabular}{ccccc}
\hline
channel: & $e^+e^-$ & $\mu^+\mu^-$ & $\tau^+\tau^-$ &  
$R_{\tau/\ell}(nS)$ \\
\hline
$\Upsilon(1S)$ & $2.38 \pm 0.11$ &  & $2.66 \pm 0.11$ & 
$0.12 \pm 0.06$\\
\hline
$\Upsilon(1S)$ &             & $2.48 \pm 0.06$ & $2.66 \pm 0.11$ & 
$0.07 \pm 0.05$\\
\hline
$\Upsilon(2S)$ & $1.92 \pm 0.17$ &    & $2.03 \pm 0.28$ & 
$0.06 \pm 0.15$ \\
\hline
$\Upsilon(2S)$ & & $1.93 \pm 0.17$ & $2.03 \pm 0.28$ & 
$0.05 \pm 0.15$ \\
\hline
$\Upsilon(3S)$ & $1.92 \pm 0.24$ &    & $2.51 \pm 0.26$ & 
$0.31 \pm 0.17$ \\
\hline
$\Upsilon(3S)$ & & $2.18 \pm 0.21$ & $2.51 \pm 0.26$ & 
$0.15 \pm 0.15$ \\
\hline
\end{tabular}
\end{center}
\end{table*}

\section{Experimental results}
In figure 1 we plot the values of $R_{\tau/\ell}(nS)$ stemming from
the results shown in Table 1. According to a hypothesis test
(where lepton universality plays the role of the null hypothesis)
lepton universality can be rejected at the $1\%$ level of
significance.

On the other hand, let us note the fact that no $\eta_b$ resonance 
has been found so
far despite intensive searches along many years. Recently
CLEO has also made a search for the
$\eta_b(1S)$ and $\eta_b(2S)$ states via hindered magnetic
dipole transitions from $\Upsilon(3S)$ with negative results too
\cite{Artuso:2004fp}. In fact, one might speculate that this failure is 
due to quite broad $^1S_0$ bottomonium states as a 
consequence of the new physics contribution \cite{Sanchis-Lozano:2005fj,
Sanchis-Lozano:2004gh}. Mixing between a CP-odd Higgs (denoted as $A^0$)
and pseudoscalar bottomonium 
states may also lead to $\Upsilon-\eta_b$ hyperfine splittings 
different from the SM expectations \cite{Sanchis-Lozano:2005fj}

\begin{figure}
\begin{center}
\includegraphics[width=23pc]{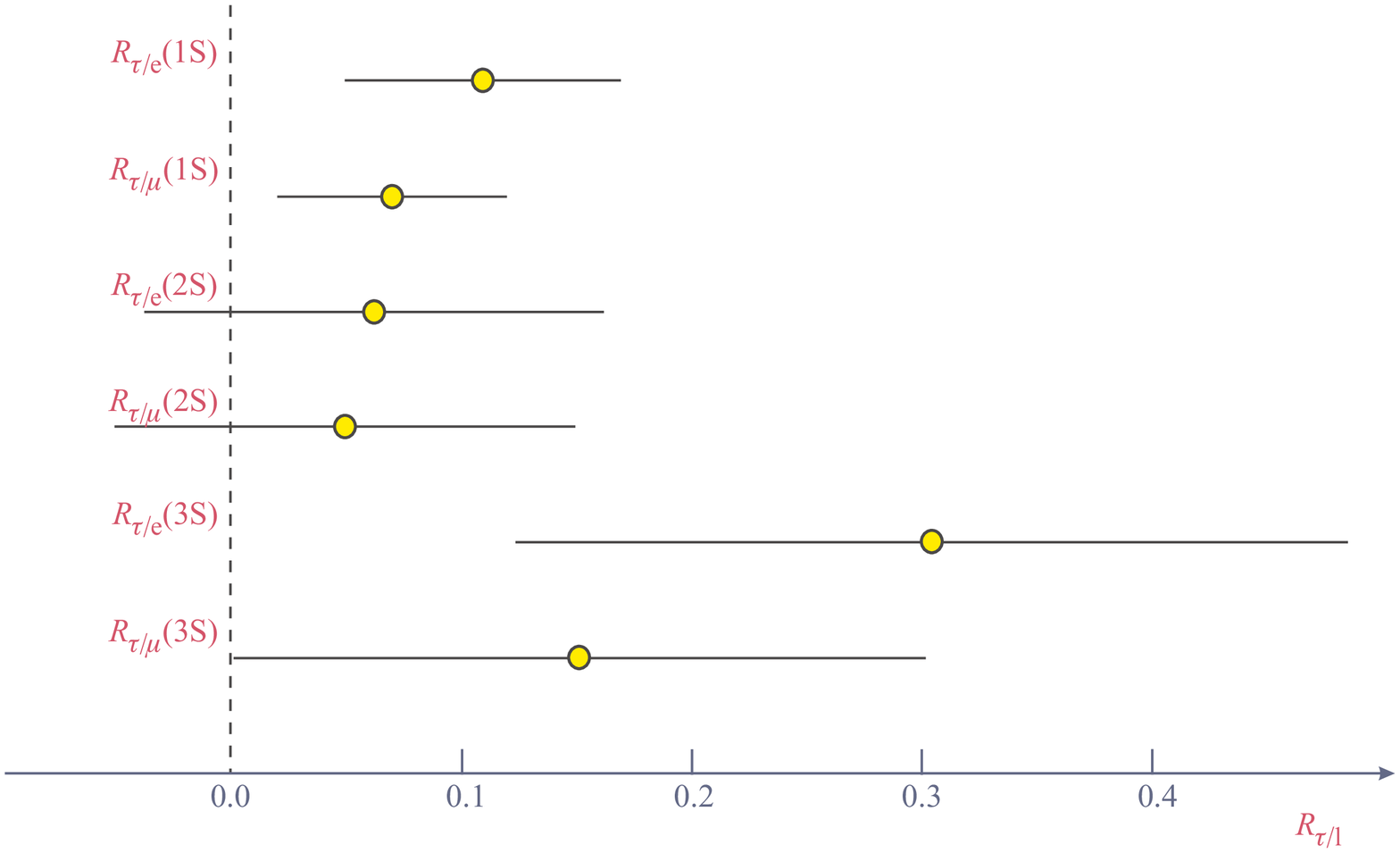}
\end{center}
\caption{Plot of $R_{\tau/\ell}$ values corresponding to Table 1.
Errors bars (partly due to systematic errors) are 
expected to be reduced after completion of the CLEO on-going analysis.}
\end{figure}
For moderate values of $\tan{\beta}$ 
(defined as the two Higgs vacuum expectation values in
a 2HDM) the $\eta_b$ decay
would be almost saturated by the Higgs-mediated annihilation into
a tau pair, according to the expression for the width 
\cite{Sanchis-Lozano:2003ha}
\begin{equation}
\Gamma[\eta_b \to \tau^+\tau^-]=\frac{3m_b^4m_{\tau}^2(1-4x_{\tau})^{1/2}\mid
R_n(0) \mid^2\tan{}^4\beta}{2\pi^2(m_{\eta_b}^2-m_{A^0}^2)^2v^4}
\end{equation}
where $x_{\tau}=m_{\tau}^2/m_{\Upsilon}^2$, $R_n(0)$ stands for
the radial $\eta_b$ wave function at the origin and $v=246$ GeV. Hence
assuming $\mathcal{B}[\eta_b \to \tau^+\tau^-]
\approx 1$,  we can conclude from Eq.(1.2) that
\begin{equation}
\mathcal{B}[\Upsilon(nS)\ \to\ \gamma_s\tau^+\tau^-] \approx
\mathcal{B}[\Upsilon(nS)\ \to\ \gamma_s\eta_b(n'S)]
\end{equation}
where the latter branching ratio can be estimated as 
a M1 transition probability given by 
\cite{Godfrey:2001eb}
\[
\mathcal{B}[\Upsilon(nS)\ \to\ \gamma_s\eta_b(n'S)]\ =\
\frac{\Gamma^{M1}}{\Gamma_\Upsilon}=\frac{16\alpha}{3}\biggl(\frac{e_b^2}{2m_b^2}\biggr)\
I\ k^3
\]
where $\alpha$ is the fine structure constant, $e_b$ and $m_b$ are the
electric charge and mass of the bottom quark, $k$ denotes the
photon energy and $I$ represents the initial and
final wave functions overlap.

Tentatively assuming universality breaking as a working hypothesis, the
results shown in Table 1 (and figure 1)  
are compatible with the following
interpretation involving new physics:
\begin{itemize}
\item There is a light CP-odd (or without definite CP) Higgs
particle whose mass lies around the $\Upsilon(1S)$.
A M1 transition of a $\Upsilon(nS)$ $(n=1,2,3)$ resonance
would yield an intermediate
$\eta_b(1S)$ state, subsequently decaying 
via a Higgs-mediated
annihilation channel into a $\tau^+\tau^-$
pair with almost unity probability - even for moderate $\tan{\beta}$ values
\item Following the previous remark, the BF's shown in Table 1 
are compatible with either an allowed
transition of the $\Upsilon(1S)$ into a $\eta_b(1S)$ state, or a
hindered transition from a $\Upsilon(2S)$ or a $\Upsilon(3S)$ into
a $\eta_b(1S)$ too, both with probabilities of
order $10^{-4}-10^{-3}$ according to potential quark model
calculations \cite{Godfrey:2001eb}. Thus, since 
$\mathcal{B}[\Upsilon(nS) \to \ell^+ \ell^-]
\simeq 2 \%$, one gets naturally
\begin{equation}
R_{\tau/\ell}(nS) \approx\ \frac{\mathcal{B}[\Upsilon(nS)\ \to\
\gamma_s\eta_b(1S)]}{\mathcal{B}[\Upsilon(nS) \to \ell \ell]}\
\approx\ 10^{-2}-10^{-1}
\end{equation}
\item In addition, one should also consider radiative decays of
$\Upsilon(2S)$ and $\Upsilon(3S)$ into an on-shell $A^0$ particle
subsequently decaying into a tauonic pair \cite{Sanchis-Lozano:2005fj}. 
Setting, e.g., 
$\tan{\beta}=15$ and $M_{\Upsilon}-M_{A^0} = 250$ MeV, one obtains
\begin{equation}
R_{\tau/\ell}(nS) \approx\ \frac{M_{\Upsilon}^2 \tan{}^2\beta}{8\pi
\alpha v^2}\biggl[1-\frac{M_{A^0}^2}{M_{\Upsilon}^2}\biggr] \approx
10^{-1}
\end{equation}

\end{itemize}

\section{Conclusions}
In the absence of any (unknown) systematic shift of the values of Table 1,
the simplest explanation of lepton universality breaking would
require a light non-standard Higgs mediating the tauonic decay
mode as advocated in this paper. Further confirmation should be
needed, e.g., by searching for those M1 soft photons in the collected
sample of tauonic decays by CLEO. I acknowledge CLEO Collaboration
for extremely valuable help when developing this work.

\end{document}